\documentclass[aps,prb,twocolumn,superscriptaddress,showpacs]{revtex4-1}

\usepackage{graphicx}
\usepackage{amsmath,amssymb}
\usepackage{bm}
\usepackage{braket}
\newcommand{\up}{\uparrow}
\newcommand{\down}{\downarrow}

\begin{document}


\title{Symmetry-protected topological phase transition in one-dimensional Kondo lattice and its realization with ultracold atoms}


\author{Masaya Nakagawa}
\email{masaya.nakagawa@riken.jp}
\affiliation{RIKEN Center for Emergent Matter Science (CEMS), Wako, Saitama 351-0198, Japan}
\author{Norio Kawakami}
\affiliation{Department of Physics, Kyoto University, Kyoto 606-8502, Japan}


\date{\today}

\begin{abstract}
We propose that ultracold alkaline-earth-like atoms confined in one-dimensional optical lattice can realize a Kondo lattice model which hosts a symmetry-protected topological (SPT) phase and an associated quantum phase transition in a controllable manner. The symmetry protection of the phase transition is discussed from two different viewpoints: topological properties related to spatial patterns of Kondo singlets, and symmetry eigenvalues of the spin states. We uncover the role of various symmetries in the phase diagram of this system by combining a weak-coupling approach by Abelian bosonization and strong-coupling pictures of ground states. Furthermore, we show that the bosonization approach correctly describes a crossover from a fermionic SPT phase to a bosonic SPT phase and an associated change of protecting symmetries as the charge degrees of freedom are frozen by the Hubbard repulsion.
\end{abstract}


\maketitle

\section{Introduction}
In the past few decades, our understanding of the role of symmetries in quantum phases has been deepened very much. Even if the ground states have the same symmetries and thus cannot be distinguished from the Landau-Ginzburg-type phase transitions with spontaneous symmetry breaking, quantum many-body systems can have numerous distinct phases. For example, the Haldane phase \cite{HaldanePhysLett, Haldane, AKLT1, AKLT2} emerging in spin-1 chains cannot be characterized by any local order parameter associated with symmetry breaking, but it possesses a non-local string order\cite{denNijs, KennedyTasaki1, KennedyTasaki2} and is still a distinct quantum phase from featureless product states. Now the Haldane phase is recognized as a typical example of symmetry-protected topological (SPT) phases.\cite{Chen, SPTbook} SPT phases are characterized by non-degenerate gapped ground states without symmetry breaking which cannot be adiabatically connected to site-product states under some symmetry constraint. Since the SPT phases can be connected to trivial product states if symmetry-breaking perturbation is allowed, the presence of symmetries is indispensable for SPT phases. In fact, the Haldane phase is distinguished from product states if either time-reversal, spatial inversion, or spin dihedral symmetry is present in the system.\cite{GuWen, Pollmann1, Pollmann2} The existence of string order is also understood from a modern perspective in connection with the symmetry protection of the Haldane phase by the spin dihedral symmetry.

The concept of SPT phases is applicable to ground states of quantum many-body systems. Hence, topological phase transitions between SPT phases (and a trivial phase) are necessarily quantum phase transitions triggered by tuning of parameters of the Hamiltonian. From this perspective, ultracold atoms with great tunability of system parameters \cite{Bloch_review} are a promising candidate for direct observation of such quantum phase transitions. For example, by engineering artificial gauge fields, transitions between topologically trivial and nontrivial band structures of non-interacting systems have been observed using fermionic\cite{Jotzu} and bosonic\cite{Aidelsburger} atoms. Since interactions can be easily introduced to atoms, an intriguing prospect in this field is realization of SPT phases with strong correlations. It potentially provides a versatile platform to study exotic phase transitions arising from the topological nature of quantum systems.

In this paper, we propose an experimentally feasible scheme to realize SPT phase transitions induced by strong interactions using ultracold fermionic atoms loaded in optical lattice. Our model is based on one of the prototypical models of strongly correlated fermions: the Kondo lattice model.\cite{Coleman_book} Using an Abelian bosonization approach, we show that a one-dimensional (1D) version of the Kondo lattice model has several distinct quantum phases including a SPT phase and identify what symmetries protect them. We demonstrate that ultracold alkaline-earth-like atoms (AEA) in optical lattices can realize the SPT phase and access the associated quantum phase transitions. 

In our setup, the phase transitions are triggered by Kondo effect which is induced by laser irradiation to the atoms using a recently proposed scheme.\cite{NakagawaKawakami} In this scheme, the laser field couples with the spin degrees of freedom of atoms and thereby realizes a tunable anisotropic spin exchange interaction. This feature enables us to engineer the quantum phase transitions with high controllability in sharp contrast to solid state realizations, where the strength of exchange interactions is intrinsic to the materials and is usually fixed. Furthermore, we show that the anisotropic exchange interaction realizes the Kondo effect with an ``unusual" spin state different from ordinary Kondo singlet. The unusual Kondo state is certainly distinct from the ordinary Kondo state by comparing their symmetry eigenvalues in terms of the spin $\pi$ rotation around $x$ or $y$ axis. Owing to this feature, we point out that the topological phase transition of this system is protected not only by its topological nature, but also by the symmetry eigenvalues of the spin states.

Besides providing the experimental setup, the other main aim of this paper is to provide a description of a crossover of SPT phases from interacting fermions to spin chains, using the bosonization approach.
In the Kondo lattice systems, interplay of mobile charges and their exchange coupling to localized spins leads to a variety of quantum phases with or without magnetic order.\cite{TsunetsuguSigristUeda, Tsvelik, FujimotoKawakami1, FujimotoKawakami2, Zachar1, LeHur, Zachar2, Garcia, PivovarovQi, FyeScalapino, Tsunetsugu, YuWhite, Shibata3, Shibata1, Shibata2, Shibata4, Peters, Silva-Valencia, Tsvelik2015, Tsvelik2016} The SPT phase that we focus on emerges in the 1D Kondo lattice with ferromagnetic exchange coupling (the double exchange model) and has been shown to approach the Haldane phase in the strong coupling limit.\cite{Tsunetsugu, Shibata3} 
However, the main difference between the Kondo lattice and the Haldane spin chains is the existence of the charge degrees of freedom. In this case, the SPT phase is no longer treated as a bosonic spin system, but must be treated as fermions. Correspondingly, when the charge fluctuations cannot be neglected, the time-reversal and spin dihedral symmetries no longer protect the Haldane phase, and only the inversion symmetry remains as the protecting symmetry. This phenomenon was previously studied using Hubbard ladders\cite{AnfusoRosch, MoudgalyaPollmann}, but here we provide alternative derivation in the present Kondo lattice setup by the bosonization method. This method transparently captures how the SPT phase composed of interacting fermions changes into the bosonic SPT phase (the Haldane phase) as the charge degrees of freedom are frozen by taking the strong coupling limit. As a result, the topological phase of the 1D Kondo lattice fits into the $\mathbb{Z}_4$ classification of interacting fermionic SPT phases protected by the inversion symmetry in addition to the charge conservation.\cite{YouXu, Shiozaki1, Shiozaki2} The fermionic aspects of the SPT phase in the present setup can be contrasted to previous studies on realization of correlated SPT phases in cold alkaline-earth atoms,\cite{Kobayashi, Kobayashi2, Nonne1, Nonne2, Nonne3, Nonne4, Nonne5, Bois, Tanimoto, Capponi} where only the strong-coupling limit and thus spin-chain models were considered.

The organization of this paper is as follows. In Sec.\ \ref{Model}, we describe our setup used in this paper and derive a 1D Kondo lattice model as an effective low-energy theory of this system. Before analyzing the Kondo lattice model, we first examine the corresponding impurity problems in Sec.\ \ref{Imp} to obtain some intuition for the problem. 
In Sec.\ \ref{Bosonization}, we proceed to an analysis of the 1D Kondo lattice model using Abelian bosonization and derive a set of renormalization group (RG) equations. Based on the RG equations, we determine the phase diagram of the system in Sec.\ \ref{Phasediagram}. In Sec.\ \ref{SymProt}, we elucidate what symmetries protect the quantum phases and describe the crossover of the SPT phase by the bosonization method. 
Finally, we conclude this paper in Sec.\ \ref{Conclusion} with discussions for experimental detections.

\section{Model\label{Model}}
We start by introducing our setup and model used in this paper. Our setup utilizes a recently proposed scheme to realize the Kondo lattice using specific properties of AEA such as $^{171}$Yb, $^{173}$Yb, and $^{87}$Sr in optical lattices. \cite{NakagawaKawakami} AEA have an electronic ground state and a long-lived excited state denoted by $^1S_0$ and $^3P_0$, respectively. We consider ultracold AEA in 1D optical lattice and assign fermionic annihilation operators of the $^1S_0$ state at lattice site $j$ to $c_{j\sigma}$, and those of the $^3P_0$ state to $f_{j\sigma}$. Here the spin indices $\sigma= -I, \cdots, I$ come from the nuclear spin degrees of freedom of atoms. Since the polarizability to light is different for each state, we can load these atoms in an optical lattice with state-dependent lattice depth. This leads to state-dependent Wannier orbitals and gives transfer integrals $t_c , t_f$ to each state. Thus the model Hamiltonian can be written in the most general form as \cite{Gorshkov, NakagawaKawakami}
\begin{align}
H=&\sum_{j, \sigma}(-t_c c_{j\sigma}^\dag c_{j+1,\sigma}-t_f f_{j\sigma}^\dag f_{j+1,\sigma}+\mathrm{h.c.})+\sum_{j,\sigma}\varepsilon_f^{(0)}n_{fj\sigma}\notag\\
&+U\sum_{j,\sigma<\sigma'} n_{cj\sigma}n_{cj\sigma'}+U_{ff}\sum_{j,\sigma<\sigma'}n_{fj\sigma}n_{fj\sigma'}\notag\\
&+U_{cf}\sum_{j,\sigma,\sigma'}n_{cj\sigma}n_{fj\sigma'}+V_{\mathrm{ex}}\sum_{j,\sigma,\sigma'}c_{j\sigma}^\dag f_{j\sigma'}^\dag c_{j\sigma'}f_{j\sigma}\notag\\
&+\sum_{j,\sigma,\sigma'}(\bm{V}\cdot\bm{\sigma}_{\sigma\sigma'}e^{i\bm{K}\cdot\bm{R}_j-i\omega t}f_{j\sigma}^\dag c_{j\sigma'}+\mathrm{h.c.}),
\label{Hamil}
\end{align}
where $n_{cj\sigma}=c_{j\sigma}^\dag c_{j\sigma}$ and $n_{fj\sigma}=f_{j\sigma}^\dag f_{j\sigma}$ count the number of particles at site $j$. $\varepsilon_f^{(0)}$ denotes the excitation energy of the $^3P_0$ state from the $^1S_0$ state. The specific values of interaction parameters $U, U_{ff}, U_{cf}, V_{\mathrm{ex}}$ depend on $s$-wave scattering lengths in corresponding collision channels and the details of optical lattice setups, namely, the Wannier-function overlaps and the trap potential for confining the atoms in one direction.\cite{Gorshkov, RZhang} Since the $s$-wave scattering lengths are independent of the nuclear spin states, the interactions possess SU($N=2I+1$) symmetry\cite{Gorshkov, Cazallila} as confirmed by experiments.\cite{Taie1, Taie2, Cappellini, Scazza, Zhang} Hereafter we assume that $U>0, U_{ff}>0$. In principle, there exist additional terms originating from a magnetic field,\cite{RZhang} but for simplicity we take the zero-field limit and avoid the complication.

The last term in Eq.\ \eqref{Hamil} is an important ingredient for our model. This term represents optical transitions between the $^1S_0$ state and the $^3P_0$ state allowed by dipole coupling with the help of hyperfine interactions.\cite{Porsev} From the Wigner-Eckart theorem, we find that the matrix elements are the inner product of a three-component vector $\bm{V}$ (which is proportional to the electric field component of the optical field) and Pauli matrices of the nuclear spin.\cite{Beri, NakagawaKawakami} $\bm{K}$ and $\omega$ are the wave number and the frequency of the optical field, respectively. We here consider a $\pi$-polarized laser field with $\bm{V}=(0,0,V)$, which does not break the time-reversal symmetry.

The explicit time dependence in the hybridization term of Eq.\ \eqref{Hamil} is eliminated by a gauge transformation $f_{j\sigma}\to e^{-i\omega t}f_{j\sigma}$. After this transformation, the energy level of the $^3P_0$ state is effectively shifted, and we replace $\varepsilon_f^{(0)}$ with $\varepsilon_f\equiv \varepsilon_f^{(0)}-\omega$. Besides the trivial time dependence due to the gauge transformation, the system is assumed to be an equilibrium state with temperature $T$ and chemical potential $\mu$. In this paper, we mainly consider the case of $T=0$ and focus on quantum phase transitions that the system exhibits.

We assume that the lattice potential is sufficiently deep for the $^3P_0$ state to suppress inelastic collisions which cause loss of atoms, and thus $t_f\ll U_{ff}$. On the other hand, the lattice potential for the $^1S_0$ state is shallow to allow the hopping between sites. To simplify the original model \eqref{Hamil}, we consider a limiting case in which the Kondo limit is achieved: $\varepsilon_f\ll \mu \ll \varepsilon_f+U_{ff}$ and $|\bm{V}|$ is sufficiently small. In this case, since the occupation number of the $^3P_0$ state in low-energy states is one at each site, we can restrict ourselves to the Hilbert subspace with $\sum_\sigma n_{fj\sigma}=1$ and derive an effective low-energy Hamiltonian using the Schrieffer-Wolff transformation.\cite{SW} The resulting low-energy theory leads to the Kondo lattice (or Kondo-Heisenberg) model
\begin{align}
H_{\mathrm{eff}}=&-t_c\sum_{j,\sigma}(c_{j\sigma}^\dag c_{j+1,\sigma}+\mathrm{h.c.})+U\sum_{j,\sigma<\sigma'} n_{cj\sigma}n_{cj\sigma'}\notag\\
&+\sum_{j,\sigma,\sigma'}(V_{\mathrm{ex}}-\sigma\sigma'J)c_{j\sigma}^\dag f_{j\sigma'}^\dag c_{j\sigma'}f_{j\sigma}\notag\\
&+J_H\sum_{j,\sigma,\sigma'}f_{j\sigma}^\dag f_{j\sigma'} f_{j+1,\sigma'}^\dag f_{j+1,\sigma}
\label{Heff}
\end{align}
where $J=2V^2(\frac{1}{|\varepsilon_f-\mu|}+\frac{1}{\varepsilon_f-\mu+U_{ff}})>0$ and $J_H=4t_f^2/U_{ff}>0$. We note that when $\sum_\sigma n_{fj\sigma}=1$, the interaction $U_{cf}$ can be incorporated into the chemical potential and therefore we omit this term from $H_{\mathrm{eff}}$.

The effective Hamiltonian \eqref{Heff} contains an effective Kondo interaction $V_{\mathrm{ex}},J$ between the two orbitals and the Heisenberg interaction $J_H$ between $^3P_0$ states. While the spin-exchanging collision $V_{\mathrm{ex}}$ is fully symmetric, the optically induced Kondo coupling $J$ breaks the spin SU($N$) symmetry due to the polarization-spin coupling in the last term in Eq.\ \eqref{Hamil}. For general $N$, this Kondo coupling is somewhat complicated, but the case of $N=2$ is simple. For $N=2$, we can rewrite the Kondo coupling as
\begin{align}
&\sum_{j,\sigma,\sigma'}(V_{\mathrm{ex}}-\sigma\sigma'J)c_{j\sigma}^\dag f_{j\sigma'}^\dag c_{j\sigma'}f_{j\sigma}\notag\\
=&-J_\perp\sum_j(S_{cj}^xS_{fj}^x+S_{cj}^yS_{fj}^y)-J_z\sum_jS_{cj}^zS_{fj}^z\notag\\
&+\mathrm{potential\; term},
\label{JK}
\end{align}
where
\begin{subequations}
\begin{align}
J_\perp&\equiv V_{\mathrm{ex}}+J/4,\label{Jperp}\\
J_z&\equiv V_{\mathrm{ex}}-J/4.\label{Jz}
\end{align}
\end{subequations}
The ``potential term" can be absorbed into the chemical potential. The spin operators are defined by $\bm{S}_{cj}=\frac{1}{2}\sum_{\sigma,\sigma'}c_{j\sigma}^\dag\bm{\sigma}_{\sigma\sigma'}c_{j\sigma'}$ and $\bm{S}_{fj}=\frac{1}{2}\sum_{\sigma,\sigma'}f_{j\sigma}^\dag\bm{\sigma}_{\sigma\sigma'}f_{j\sigma'}$, where $\bm{\sigma}$ is the three-component Pauli matrices. This interaction is just an anisotropic XXZ-type exchange coupling between the $^1S_0$ and the $^3P_0$ states.

Hereafter, we analyze the low-energy effective model \eqref{Heff} for $N=2$. Experimentally, this case can be realized using two specific spin states $\sigma$ and $-\sigma$ selected from $2I+1$ nuclear spins of AEA.

\section{Kondo impurity\label{Imp}}
\begin{figure}[t]
\includegraphics[width=8.5cm]{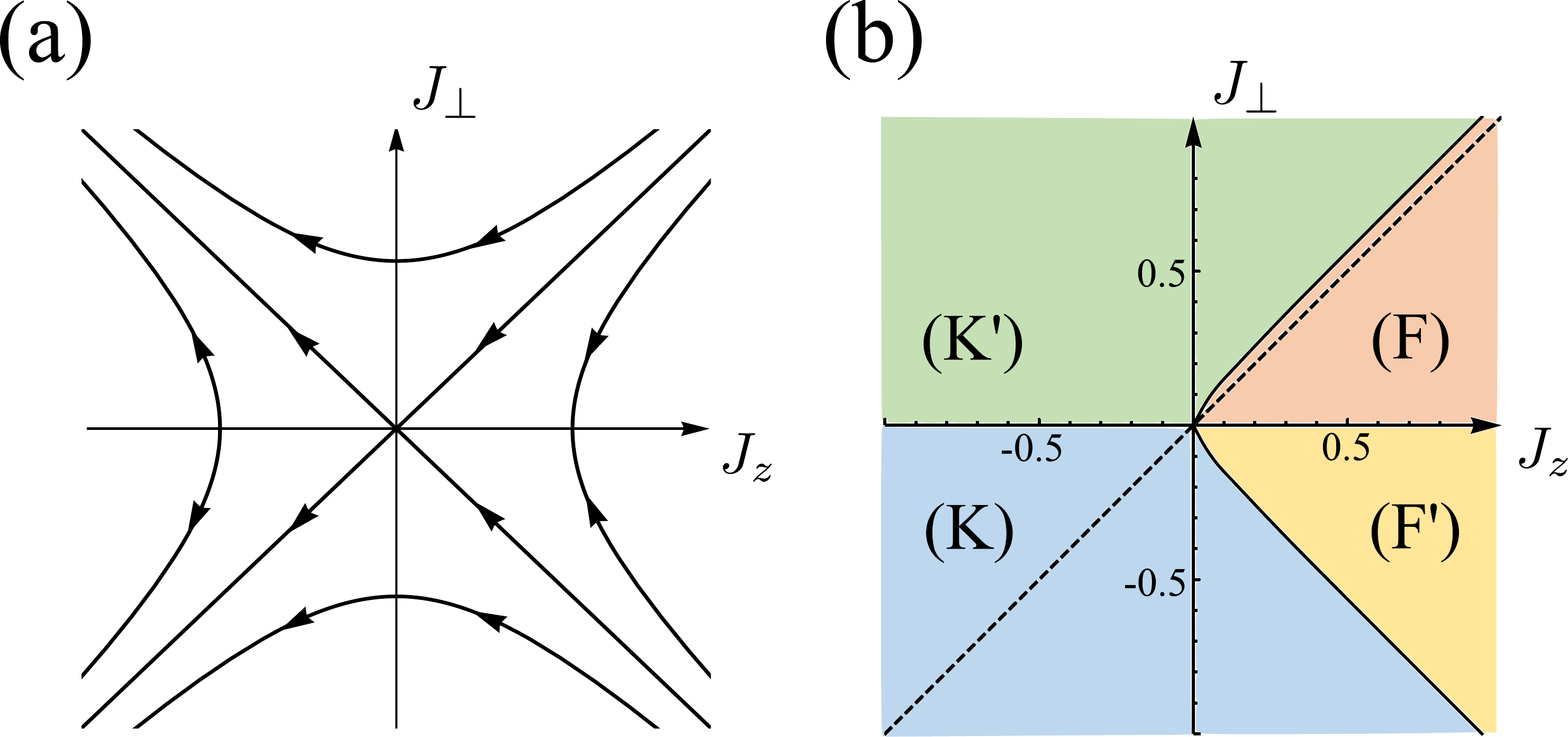}
\caption{(a) RG flow for the Kondo impurity in 3D. (b) Phase diagram of the Kondo impurity in 1D. We set $J_{\perp F}=J_{\perp B}=J_\perp, J_{zF}=J_{zB}=J_z$, $v_F=1$, and $g_2=0.5$. The broken line indicates the isotropic line on which $J_\perp=J_z$ is satisfied.}
\label{fig_imp}
\end{figure}

Before studying the full Kondo lattice Hamiltonian \eqref{Heff}, it is helpful to gain some insights from what happens when a single atom in the $^3P_0$ state is immersed into the Fermi sea of $^1S_0$ atoms as an impurity. Here we summarize known basic results\cite{Coleman_book, Anderson, LeeToner, FurusakiNagaosa} and extend them to obtain a phase diagram in Fig.\ \ref{fig_imp} (b) which is important for later analysis. Let us consider the following Kondo impurity problems:
\begin{align}
H_{3D}=-t_c&\sum_{\langle i,j\rangle,\sigma}(c_{i\sigma}^\dag c_{j\sigma}+\mathrm{h.c.})\notag\\
&-J_\perp(S_{c0}^xS_{\mathrm{imp}}^x+S_{c0}^yS_{\mathrm{imp}}^y)-J_zS_{c0}^zS_{\mathrm{imp}}^z,\\
H_{1D}=-t_c&\sum_{j,\sigma}(c_{j,\sigma}^\dag c_{j+1,\sigma}+\mathrm{h.c.})+U\sum_j n_{cj\up}n_{cj\down}\notag\\
&-J_\perp(S_{c0}^xS_{\mathrm{imp}}^x+S_{c0}^yS_{\mathrm{imp}}^y)-J_zS_{c0}^zS_{\mathrm{imp}}^z.\label{H1D}
\end{align}
In both cases, a single impurity spin is located at $j=0$. The impurity interacts with itinerant fermions living in 3D (or 1D) lattices via anisotropic Kondo couplings. In the 1D case, we have introduced the interaction between itinerant fermions and consider a metallic Tomonaga-Luttinger-liquid region away from half filling. If we set a high-energy cutoff (the bandwidth) as $D$, the RG equations for the 3D case are\cite{Anderson}
\begin{subequations}
\begin{align}
\frac{dJ_\perp}{d\ell}&=-\rho_0 J_\perp J_z,\\
\frac{dJ_z}{d\ell}&=-\rho_0J_\perp^2,
\end{align}
\end{subequations}
where $d\ell=-d\ln D$. Here $\rho_0$ is the density of states at the Fermi energy. The flow diagram is depicted in Fig.\ \ref{fig_imp} (a). The system has two fixed points characterized by growth of Kondo coupling with different signs of $J_\perp$. The fixed point with $J_\perp\to-\infty, J_z\to-\infty$ corresponds to the ordinary Kondo effect with isotropic antiferromagnetic interactions. However, an important aspect arises from the other fixed point in Fig. \ref{fig_imp} (a) for the present setup in cold atoms. As found from Eqs.\ \eqref{Jperp} and \eqref{Jz}, when the laser-induced Kondo coupling is sufficiently strong, we reach the fixed point with $J_\perp\to\infty, J_z\to-\infty$. The nature of this fixed point can be extracted from a transformation
\begin{equation}
(S_{\mathrm{imp}}^x, S_{\mathrm{imp}}^y, S_{\mathrm{imp}}^z)\rightarrow(-S_{\mathrm{imp}}^x, -S_{\mathrm{imp}}^y, S_{\mathrm{imp}}^z),
\label{anomKondotfm}
\end{equation}
which is equivalent to flipping the sign of $J_\perp$. Note that this transformation keeps the commutation relation intact. Since the singlet state $\ket{\down}_c\ket{\up}_f-\ket{\up}_c\ket{\down}_f$ is transformed into $\ket{\down}_c\ket{\up}_f+\ket{\up}_c\ket{\down}_f$ by this procedure, we find that the fixed point describes the Kondo effect with Kondo ``singlet" $\ket{\down}_c\ket{\up}_f+\ket{\up}_c\ket{\down}_f$. 

The 1D case was studied by Refs.\ \onlinecite{LeeToner, FurusakiNagaosa}, and the situation is somewhat different from 3D. In 1D, the forward scattering off the impurity and the backward one are distinguished. Hence we must double the coupling constants for the Kondo coupling: $J_{\perp F}, J_{\perp B}, J_{zF}, J_{zB}$ where the subscript $F$ ($B$) denotes the forward (backward) process. Then the RG equations are given by
\begin{subequations}
\begin{align}
\frac{dJ_{\perp F}}{d\ell}&=-\frac{1}{2\pi v_F}(J_{\perp F}J_{zF}+J_{\perp B}J_{zB}),\label{impRG1D_1}\\
\frac{dJ_{\perp B}}{d\ell}&=-\frac{1}{2\pi v_F}(J_{\perp F}^2+J_{\perp B}^2),\\
\frac{dJ_{zF}}{d\ell}&=\frac{1}{2\pi v_F}(g_2 J_{\perp B}-J_{\perp F}J_{zB}-J_{\perp B}J_{zF}),\\
\frac{dJ_{zB}}{d\ell}&=\frac{1}{2\pi v_F}(g_2 J_{zB}-2J_{\perp F}J_{\perp B}),\label{impRG1D_4}
\end{align}
\end{subequations}
where $g_2$ denotes the matrix element of the forward scattering process between itinerant fermions due to the Hubbard repulsion in Eq.\ \eqref{H1D}. $v_F$ is the Fermi velocity. 
By integrating Eqs.\ \eqref{impRG1D_1} - \eqref{impRG1D_4} numerically, we obtain a phase diagram in Fig.\ \ref{fig_imp} (b), although the flow diagram was shown only for the isotropic ($J_\perp=J_z$) case in Ref.\ \onlinecite{FurusakiNagaosa}. The phase (K) shows the ordinary Kondo effect and the phase (K') shows the ``unusual" Kondo effect as in the 3D case. A peculiar point in 1D is the existence of a new phase (F) where the exchange coupling grows to strong coupling starting from bare \textit{ferromagnetic} interactions. This fixed point appears only when $g_2>0$ is included,\cite{FurusakiNagaosa} and therefore we need to consider the Hubbard repulsion in Eq.\ \eqref{H1D}. At the fixed point, the coupling constants grow as $J_{\perp F}\to-\infty, J_{\perp B}\to\infty, J_{zF}\to-\infty, J_{zB}\to\infty$. Note that the signs are negative for the forward processes and positive for the backward ones. From this observation, it turns out that the fixed point describes growth of nearest-neighbor antiferromagnetic Kondo coupling which leads to a Kondo singlet state with the adjacent sites of the impurity, while the onsite Kondo coupling is kept finite.\cite{FurusakiNagaosa} The phase (F') is not important for the later discussions, but the nature of this phase is also understood by the transformation \eqref{anomKondotfm}. In the subsequent sections, we show that the phase diagram of 1D Kondo lattice has similarity to the 1D impurity case.

\section{Renormalization group analysis of 1D anisotropic Kondo lattice\label{Bosonization}}
Let us now proceed to the analysis of the 1D Kondo lattice model. Hereafter we consider the Hamiltonian \eqref{Heff} for $N=2$ with the half-filling condition for $^1S_0$ states. To analyze the low-energy behavior of the system, we apply Abelian bosonization\cite{Giamarchi} to the Hamiltonian using the following identity: 
\begin{align}
c_{j\sigma}=\frac{1}{\sqrt{2\pi}}(&\eta_{R\sigma}e^{ik_Fx}e^{i(\theta_{1\sigma}(x)-\phi_{1\sigma}(x))}\notag\\
&+\eta_{L\sigma}e^{-ik_Fx}e^{i(\theta_{1\sigma}(x)+\phi_{1\sigma}(x))})
\end{align}
where $x=ja$ is the continuum space variable and the boson fields $\phi, \theta$ satisfy a commutation relation $[\phi_{1\sigma}(x),\nabla\theta_{1\sigma'}(y)]=i\pi\delta_{\sigma\sigma'}\delta(x-y)$. In the above expression, the boson field $\phi$ is compactified as $\phi\sim\phi+2\pi$. 
The Fermi momentum $k_F$ is fixed at $k_F=\pi/2a$ due to the half-filling condition. $\eta_{R/L\sigma}$ is a Klein factor expressed in terms of Majorana fermions satisfying $\{\eta_{\alpha},\eta_{\beta}\}=2\delta_{\alpha\beta}$, which ensures the anticommutation relation between the right mover and the left mover. Similarly, we introduce the boson fields $\phi_{2\sigma}, \theta_{2\sigma}$ for the $f_{j\sigma}$ fermions. Following standard calculations detailed in Appendix \ref{App_bos}, we obtain
\begin{widetext}
\begin{align}
H_{\mathrm{eff}}=&H_0+H_{\mathrm{int}},\label{H_bos}\\
H_0=&\frac{1}{2\pi}\int dx(u_{1c}K_{1c}(\nabla\theta_{1c})^2+\frac{u_{1c}}{K_{1c}}(\nabla\phi_{1c})^2)
+\sum_{\nu=\pm}\frac{1}{2\pi}\int dx(u_{\nu}K_{\nu}(\nabla\theta_{\nu})^2+\frac{u_{\nu}}{K_{\nu}}(\nabla\phi_{\nu})^2),\\
H_{\mathrm{int}}=&g_U\int dx \cos(2\sqrt{2}\phi_{1c})
-g_{K\perp F+}
\int dx\cos 2\phi_+\cos 2\theta_-
-g_{K\perp F-}
\int dx\cos 2\phi_-\cos 2\theta_-\notag\\
&-g_{K\perp B}
\int dx\sin\sqrt{2}\phi_{1c}\cos 2\theta_-
-g_{KzB+}
\int dx\sin\sqrt{2}\phi_{1c}\cos 2\phi_+
-g_{KzB-}
\int dx\sin\sqrt{2}\phi_{1c}\cos 2\phi_-,
\label{Hint}
\end{align}
\end{widetext}
where
\begin{subequations}
\begin{gather}
u_{1c}=2t_c a\sqrt{1+\frac{U}{2\pi t_c}},\\
u_\pm=2t_c a\sqrt{\bigl(1-\frac{U}{2\pi t_c}\bigr)\bigl(1\mp\frac{\alpha J_z}{2\pi u}\bigr)},\\
K_{1c}=1/\sqrt{1+\frac{U}{2\pi t_c}},\\
K_\pm=\frac{1}{\sqrt{1\mp\frac{\alpha J_z}{2\pi u}}},
\end{gather}
\end{subequations}
and the coupling constants are
\begin{subequations}
\begin{gather}
g_{U}=\frac{U}{2\pi^2\alpha},\\
g_{K\perp F+}=g_{K\perp F-}=\frac{1}{2m}g_{K\perp B}=\frac{J_\perp}{2\pi^2\alpha},\\
g_{KzB+}=g_{KzB-}=\frac{mJ_z}{2\pi^2\alpha}.
\end{gather}
\end{subequations}
Here $\alpha$ denotes the short-range cutoff and $m=\langle \sin\sqrt{2}\phi_{2c}\rangle$ is the expectation value of the gapped charge mode of localized $f$ fermions. The new boson fields for the charge mode (of $^1S_0$ state) $\phi_{1c},\theta_{1c}$ and the total/relative spin modes $\phi_\pm, \theta_\pm$ are defined as
\begin{subequations}
\begin{align}
\phi_{1c}\equiv&\frac{1}{\sqrt{2}}(\phi_{1\up}+\phi_{1\down}),\\
\theta_{1c}\equiv&\frac{1}{\sqrt{2}}(\theta_{1\up}+\theta_{1\down}),\\
\phi_\pm\equiv&\frac{1}{2}(\phi_{1\up}-\phi_{1\down}\pm(\phi_{2\up}-\phi_{2\down})),\\
\theta_\pm\equiv&\frac{1}{2}(\theta_{1\up}-\theta_{1\down}\pm(\theta_{2\up}-\theta_{2\down})).
\end{align}
\end{subequations}
For later convenience, we name each term in Eq.\ \eqref{Hint} as $H_U, H_{K\perp F+},H_{K\perp F-}, H_{K\perp B}, H_{KzB+},$ and $H_{KzB-}$, where the subscripts correspond to those of the coupling constants (see Appendix \ref{App_bos}).

The low-energy behavior of the model \eqref{H_bos} is deduced from perturbative RG analysis in terms of $H_{\mathrm{int}}$. Since the unperturbed theory $H_0$ is free bosons and thus is a conformal field theory (CFT), the RG equations can be derived from the CFT data of the free boson theory, i.e. scaling dimensions and operator-product-expansion coefficients.\cite{Fradkin} After some calculations, we arrive at a set of RG equations when the cutoff is changed from $\alpha$ to $e^{d\ell}\alpha$, as
\begin{subequations}
\begin{align}
\frac{dK_{1c}}{d\ell}=&-K_{1c}^2(2\tilde{g}_{U}+2\tilde{g}_{K\perp B}^2+\tilde{g}_{KzB+}^2+\tilde{g}_{KzB-}^2),\label{RGK1c}\\
\frac{dK_{+}}{d\ell}=&-K_+^2(2\tilde{g}_{K\perp F+}^2+2\tilde{g}_{KzB+}^2),\\
\frac{dK_{-}}{d\ell}=&-K_-^2(2\tilde{g}_{K\perp F-}^2+2\tilde{g}_{KzB-}^2)\notag\\
&+2\tilde{g}_{K\perp F+}^2+2\tilde{g}_{K\perp F-}^2+4\tilde{g}_{K\perp B}^2,
\end{align}
and
\begin{align}
\frac{d\tilde{g}_{U}}{d\ell}=&(2-2K_{1c})\tilde{g}_{U}+\tilde{g}_{K\perp B}^2+\tilde{g}_{KzB+}^2+\tilde{g}_{KzB-}^2,\label{RGU}\\
\frac{d\tilde{g}_{K\perp F+}}{d\ell}=&(2-K_+-\frac{1}{K_-})\tilde{g}_{K\perp F+}-\tilde{g}_{K\perp B}\tilde{g}_{KzB+},\label{RGKperpFp}\\
\frac{d\tilde{g}_{K\perp F-}}{d\ell}=&(2-K_--\frac{1}{K_-})\tilde{g}_{K\perp F-}-\tilde{g}_{K\perp B}\tilde{g}_{KzB-},\\
\frac{d\tilde{g}_{K\perp B}}{d\ell}=&(2-\frac{1}{2}K_{1c}-\frac{1}{K_-})\tilde{g}_{K\perp B}\notag\\
&-\tilde{g}_{K\perp F+}\tilde{g}_{KzB+}-\tilde{g}_{K\perp F-}\tilde{g}_{KzB-}+\frac{1}{2}\tilde{g}_{U}\tilde{g}_{K\perp B},\\
\frac{d\tilde{g}_{KzB+}}{d\ell}=&(2-\frac{1}{2}K_{1c}-K_+)\tilde{g}_{KzB+}\notag\\
&-\tilde{g}_{K\perp F+}\tilde{g}_{K\perp B}+\frac{1}{2}\tilde{g}_{U}\tilde{g}_{KzB+}\label{RGKz2p},\\
\frac{d\tilde{g}_{KzB-}}{d\ell}=&(2-\frac{1}{2}K_{1c}-K_-)\tilde{g}_{KzB-}\notag\\
&-\tilde{g}_{K\perp F-}\tilde{g}_{K\perp B}+\frac{1}{2}\tilde{g}_{U}\tilde{g}_{KzB-},\label{RGKz2m}
\end{align}
\end{subequations}
up to the second order perturbation theory. Here the dimensionless coupling constants are defined by $\tilde{g}_{\alpha}\equiv \frac{1}{\pi}g_{\alpha}a^{2-\Delta_\alpha}$, where $\Delta_\alpha$ is the scaling dimension of the perturbation.

\section{Phase diagram\label{Phasediagram}}
\begin{figure}[h]
\includegraphics[width=5.5cm]{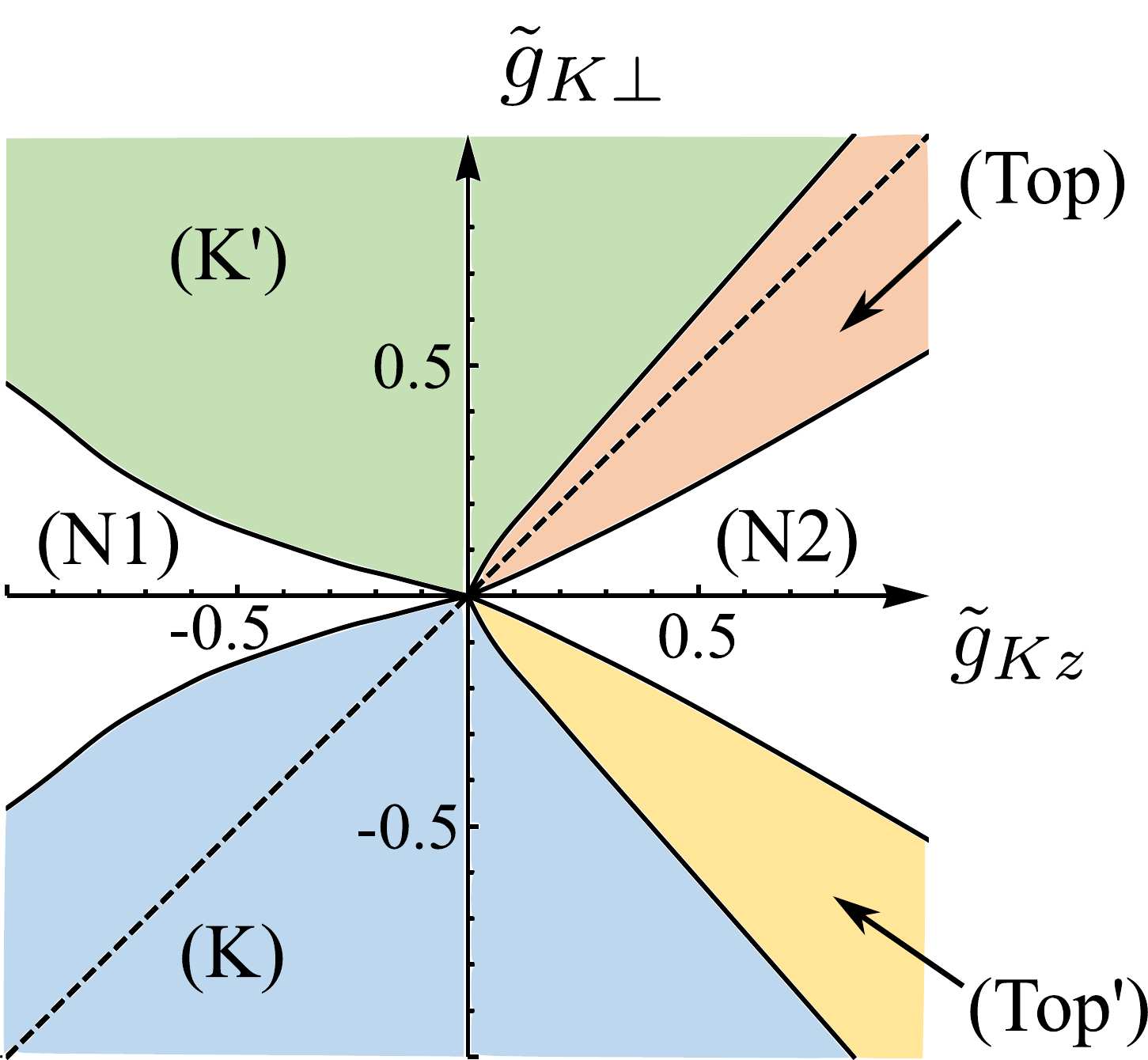}
\caption{Phase diagram of the 1D anisotropic Kondo lattice model. The broken line indicates the isotropic line on which $J_\perp=J_z$ is satisfied.}
\label{fig_phase}
\end{figure}

The zero-temperature phase diagram of the system is determined by fixed points derived from the RG equations \eqref{RGK1c} - \eqref{RGKz2m}. Numerical solutions of the RG equations indicate the phase diagram summarized in Fig.\ \ref{fig_phase}. In calculating Fig.\ \ref{fig_phase}, we have set $\tilde{g}_U=0.1$ and the initial values of the coupling constants as $\tilde{g}_{K\perp F\pm}=\frac{1}{2}\tilde{g}_{K\perp B}=\tilde{g}_{K\perp}$ and $\tilde{g}_{KzB\pm}=\tilde{g}_{Kz}$. The phase diagram is fully symmetric with respect to the sign of $\tilde{g}_{K\perp}$. As seen from scaling dimensions, the low-energy behavior is mainly governed by relevant terms $H_{K\perp B}, H_{KzB+}$, and $H_{KzB-}$. Each phase is characterized by the most divergent interactions as follows:
\begin{center}
\begin{table}[h]
\begin{tabular}{lll}
(K) & $\tilde{g}_{K\perp B}\to-\infty,$ & $\tilde{g}_{KzB+}\to-\infty$ \\
(K') & $\tilde{g}_{K\perp B}\to+\infty,$ & $\tilde{g}_{KzB+}\to-\infty$ \\
(Top) & $\tilde{g}_{K\perp B}\to+\infty,$ & $\tilde{g}_{KzB+}\to+\infty$ \\
(Top') & $\tilde{g}_{K\perp B}\to-\infty,$ & $\tilde{g}_{KzB+}\to+\infty$ \\
(N1) & $\tilde{g}_{KzB+}\to-\infty,$ & $\tilde{g}_{KzB-}\to-\infty$\\
(N2) & $\tilde{g}_{KzB+}\to+\infty,$ & $\tilde{g}_{KzB-}\to+\infty$
\end{tabular}
\end{table}
\end{center}
The phase boundary between (K') and (Top) [or (K) and (Top')] is signaled by the change of the sign of $\tilde{g}_{KzB+}$. On the other hand, the transitions to the phase (N1) or (N2) are determined by competition between $H_{K\perp B}$ and $H_{KzB-}$, which cannot be minimized simultaneously. Since the renormalization is stopped around $\tilde{g}(\ell)\sim 1$, we determine those phase boundaries by examining which of $\tilde{g}_{K\perp B}$ and $\tilde{g}_{KzB-}$ first grows to unity. We note that the role of less relevant $H_U, H_{K\perp F\pm}$ terms is the shift of phase boundaries. If we truncate the RG equations up to the tree level, the phase boundary between the phase (K') and the phase (Top) is located at $\tilde{g}_{Kz}=0$. Thus the generation of effective couplings due to less relevant interactions significantly shifts the phase boundaries. We note that the precise positions of phase boundaries depend on the Luttinger parameter.

Qualitatively, our weak-coupling calculation by the perturbative RG approach reproduces the phase diagram of 1D anisotropic Kondo lattice obtained by strong coupling expansion and exact diagonalization of a small cluster.\cite{Shibata3} Although Ref.\ \onlinecite{Shibata3} explained each phase based on spin-chain pictures in the strong coupling limit, we here point out that the phase diagram has some resemblance with the impurity phase diagram in Fig.\ \ref{fig_imp} (b) except for the appearance of the phases (N1) and (N2) which denote N\'{e}el orders. 
This resemblance can be understood to some extent by comparing the RG equations \eqref{impRG1D_1}-\eqref{impRG1D_4} and \eqref{RGKperpFp}-\eqref{RGKz2m}. Hence, our weak-coupling approach provides a complementary understanding of the phase diagram in Ref.\ \onlinecite{Shibata3}. In the following subsections, we explain the details of each phase, keeping in mind the connection to the impurity physics.

\begin{figure}
\includegraphics[width=7.5cm]{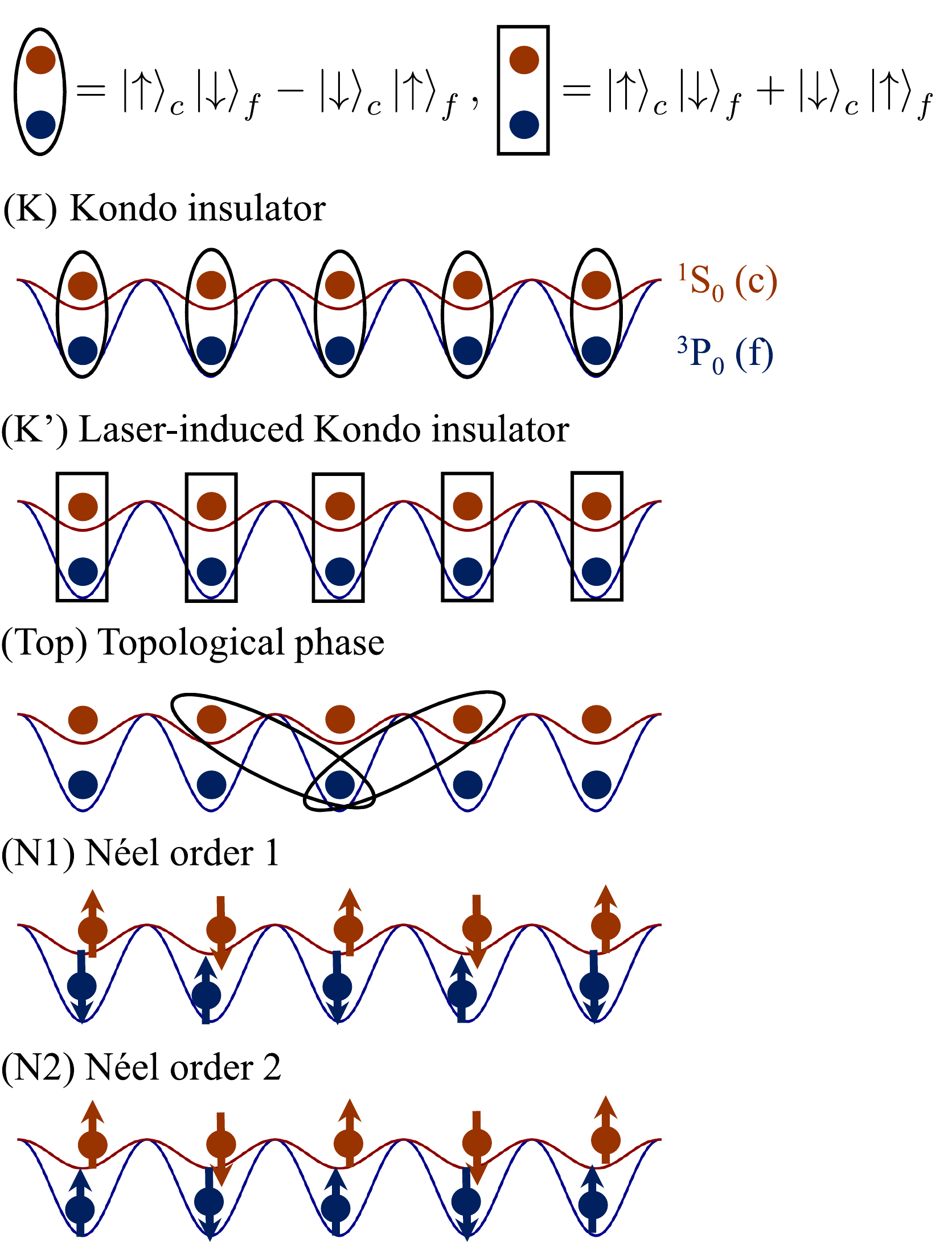}
\caption{Schematic pictures of the phases of the 1D Kondo lattice. The red (blue) balls illustrate atoms in the $^1S_0$ ($^3P_0$) state loaded in a shallow (deep) optical lattice potential. In the figure of the phase (Top), the singlet formation is represented by the central site for clarity of illustration.}
\label{fig_state}
\end{figure}

\subsection{Kondo insulator}
The phases (K), (K'), (Top), and (Top') are described by pinning of $\phi_{1c}, \phi_+$ and $\theta_-$ to their potential minimum, leading to disordered ground states with an energy gap. The phase (K) corresponds to the growth of on-site antiferrromagnetic Kondo coupling, which means the formation of the Kondo insulator.\cite{TsunetsuguSigristUeda} The strong coupling picture of this phase is illustrated in Fig.\ \ref{fig_state}, where the Kondo singlet at each site opens the energy gaps in charge and spin sectors. 
We note that the Kondo coupling effectively generates the Hubbard repulsion between conductive fermions due to Eq.\ \eqref{RGU}. Hence, even if the bare Hubbard interaction is switched off, the Kondo insulator cannot be distinguished from the Mott insulating state at least in the low-energy region. At the strong coupling limit, the Kondo insulating state approaches to the rung-singlet state if we regard the system as a spin-1/2 ladder. 

\subsection{Laser-induced Kondo insulator}
With sufficiently strong laser coupling, the phase (K') is realized owing to Eqs.\ \eqref{Jperp} and \eqref{Jz}. This phase is also a Kondo insulator, but is composed of the ``unusual" Kondo effect described in Sec.\ \ref{Imp} by a strong coupling fixed point with anisotropic Kondo coupling. As in Sec.\ \ref{Imp}, a physical picture of this Kondo insulator is obtained by a unitary transformation
\begin{equation}
f_{j\sigma}\to \mathrm{sgn}(\sigma)f_{j\sigma},
\label{lattice_AK_tfm}
\end{equation}
which flips the sign of $S_{fj}^x, S_{fj}^y$ and maps the Kondo singlet $\ket{\down}_c\ket{\up}_f-\ket{\up}_c\ket{\down}_f$ to $\ket{\down}_c\ket{\up}_f+\ket{\up}_c\ket{\down}_f$. Thus, in the strong coupling limit, the phase (K') is described by an insulating state where the $^1S_0$ state and the $^3P_0$ state form the ``Kondo singlet" $\ket{\down}_c\ket{\up}_f+\ket{\up}_c\ket{\down}_f$ at each site (Fig. \ref{fig_state}). The unusual Kondo singlet has total spin 1 with $S_c^z+S_f^z=0$, and therefore the expectation value of total spin is nonzero in the $x, y$ plane: $\langle(S_{cj}^x+S_{fj}^x)^2+(S_{cj}^y+S_{fj}^y)^2\rangle\neq0$. In the language of spin systems, this phase is very similar to the so-called large-D phase\cite{Shibata3} where a strong single-ion anisotropy favors the $S^z=0$ state in spin-1 systems.\cite{ChenHidaSanctuary}  

\subsection{Topological phase}
The phase (Top) in Fig.\ \ref{fig_phase} is a nontrivial topological phase protected by the spatial inversion symmetry, whose topological aspects are described in the next section. This phase includes the case of isotropic ferromagnetic Kondo coupling indicated by the broken line in Fig.\ \ref{fig_phase}. This phase is smoothly connected to the Haldane phase in spin ladders\cite{StrongMillis1, StrongMillis2, Shelton, Lecheminant} in the strong coupling limit $U\to\infty$ (or $J_\perp, J_z\to\infty$).\cite{Tsunetsugu} An intuitive picture of this fixed point can be obtained by considering a nearest-neighbor Kondo coupling
\begin{equation}
\tilde{H}_K\equiv -\tilde{J}\sum_j(\bm{S}_{c,j-1}+\bm{S}_{c,j+1})\cdot\bm{S}_{f,j}
\end{equation}
in addition to the original on-site Kondo coupling. The bosonized Hamiltonian is changed as
\begin{subequations}
\begin{align}
H_{K\perp F}&\to \frac{J_\perp+\tilde{J}}{J_\perp}H_{K\perp F},\\
H_{K\perp B}&\to \frac{J_\perp-\tilde{J}}{J_\perp}H_{K\perp B},\\
H_{Kz F}&\to \frac{J_z+\tilde{J}}{J_z}H_{Kz F},\\
H_{Kz B}&\to \frac{J_z-\tilde{J}}{J_z}H_{Kz B}.
\end{align}
\end{subequations}
Thus, the fixed point is equivalent to growth of the \textit{nearest-neighbor antiferromagnetic} Kondo coupling, similarly to the phase (F) appearing in the 1D Kondo impurity problem in Sec.\ \ref{Imp}, while the on-site Kondo coupling is ferromagnetic and kept finite. An intuitive picture is illustrated in Fig.\ \ref{fig_state}. The formation of the non-local Kondo singlets is reminiscent of 1D topological Kondo insulators\cite{Alexandrov, Lobos1, Lobos2, Hagymasi} realized by a $p$-wave Kondo coupling. In fact, the low-energy effective theory is the same as that of the 1D topological Kondo insulators.\cite{Lobos1} 

We note that the nature of the phase (Top') is related to the topological phase (Top) via the transformation \eqref{lattice_AK_tfm}, although this phase cannot be realized because the coupling constants cannot be manipulated into the corresponding parameter region, since $J$ is always positive in Eqs.\ \eqref{Jperp} and \eqref{Jz}.

\subsection{N\'{e}el order}
The phases (N1) and (N2) which appear near the ``Ising line" $J_\perp=0$ have an antiferromagnetic N\'{e}el order with spontaneously broken spin flip symmetry. The ordered spin patterns are illustrated in Fig.\ \ref{fig_state}. To understand the appearance of the N\'{e}el order, it is useful to consider the case of $J_\perp=0$. In this case, the remaining perturbation terms are $H_{U}, H_{KzB+},$ and $H_{KzB-}$, which are all relevant for $U>0$ and thus lock the fields $\phi_{1c}, \phi_{+}, \phi_{-}$ at their potential minimum. The locking of $\phi_\pm$ leads to the nonzero expectation value of $N_{c,f}^z(x)$ [Eq.\ \eqref{Ncz}], implying the emergence of the N\'{e}el ordering. Since the pinning of $\phi_{1c}, \phi_{+}, \phi_{-}$ opens the energy gap and the gap cannot be collapsed by infinitesimal perturbation, the N\'{e}el order should persist to some threshold value of $J_\perp$. However, the threshold value should not exceed $|J_z|$, since at the isotropic line $|J_{\perp}|=|J_z|$ we obtain the Kondo insulating phases or the topological phases by non-Abelian bosonization.\cite{FujimotoKawakami1, FujimotoKawakami2}

The existence of the N\'{e}el order can also be naturally understood from the corresponding impurity problem. When the Kondo coupling is completely Ising-like with vanishing $J_\perp$, we do not have the Kondo effect and the impurity ground state is doubly degenerate where the spins of conduction electrons and the impurity align ferromagnetically in $J_z>0$ and antiferromagnetically in $J_z<0$. Thus the residual impurity entropy $\ln 2$ should be washed out by spin ordering in the case of Kondo lattice systems.

\section{Symmetry protection\label{SymProt}}

All the quantum phases of the 1D anisotropic Kondo lattice described in Sec.\ \ref{Phasediagram} have energy gaps both in charge and spin excitations. While the N\'{e}el orders can be characterized by spontaneous breaking of the spin flip symmetry, rest four phases have the same symmetries and cannot be characterized by spontaneous symmetry breaking. In this section, we describe the roles of various symmetries in the system and provide conditions to distinguish these four phases as different quantum phases.

\subsection{Protection by spatial inversion symmetry: a crossover from a fermionic SPT phase to a bosonic SPT phase}
First, we describe what symmetry protects the topological phase (Top). The topological phase approaches the Haldane phase in spin chains in the strong coupling limit $U\to\infty$. Hence the topological phase of the 1D Kondo lattice is expected to be stable under either time-reversal, spatial inversion, or spin dihedral symmetry, if $U$ is sufficiently large and the charge degrees of freedom are frozen in the low-energy part of the Hilbert space. However, if $J_\perp, J_z$ and $U$ are small compared to the kinetic energy $t_c$, we can no more regard the system as bosonic (spin) systems and must treat it as interacting fermions. It was previously shown\cite{AnfusoRosch, MoudgalyaPollmann} that the Haldane phase with mobile charge degrees freedom is unstable and can be adiabatically connected to a trivial band insulator by only breaking inversion symmetry, even if the time-reversal and spin rotation symmetries are preserved. This fact stems from that the charge fluctuations in the low-energy Hilbert space mix the integer-spin representation of the original spin chain and that of half-odd-integer spin, invalidating the proof of the symmetry protection of the Haldane phase. Hence the only protecting symmetry of the topological phase is the inversion symmetry. Under the inversion symmetry, the degeneracy of the entanglement spectrum, which is a fingerprint of the SPT phase, still persists.\cite{MoudgalyaPollmann} A similar degenerate structure of the entanglement spectrum is also observed in 1D topological Kondo insulator\cite{Hagymasi} and 1D periodic Anderson model with Hund coupling,\cite{Hagymasi2} indicating the existence of the SPT phase. A related study on a three-leg Hubbard ladder has been also performed.\cite{Nourse}

Here we show that the above difference between the fermionic and the bosonic SPT phases is captured by the bosonization method in the present Kondo lattice system. To apply the symmetry protection argument to the present Kondo lattice system, we summarize the symmetry transformation of bosonized fields for each symmetry of the system in Table \ref{SymTfm}. Let us first consider the strong coupling limit $U\to\infty$. In this case, the charge mode $\phi_{1c}$ is completely frozen to the potential minimum of the Umklapp term $H_U$. The remaining degrees of freedom are the total and relative spin modes $\phi_\pm, \theta_\pm$, and they are equivalent to the effective theory of the corresponding spin ladder system.\cite{Shelton, Lecheminant} Hence the proof of symmetry protection can be performed in parallel with the case of the spin ladder\cite{Fuji} (see Appendix \ref{App_SPT} for the description of SPT phases by bosonization).
The gapped phases are characterized by the expectation values of the boson fields $\phi_+$ and $\theta_-$. 
To connect the topological phase with the trivial phases, a shift of the expectation value of $\phi_+$ must take place, which breaks the time-reversal, spatial inversion, and spin dihedral symmetries. Hence the topological phase is protected by those three symmetries. However, the situation is changed if we consider a weakly interacting regime. If the Hubbard interaction $U$ is sufficiently small, the Umklapp term is less relevant than the Kondo couplings $H_{K\perp B}, H_{KzB+}$, and $H_{KzB-}$. Thus the low-energy behavior is mainly governed by the Kondo couplings. In this case, we can adiabatically connect the topological phase (Top) and the ordinary Kondo insulator (K) without closing the energy gap, by shifting the expectation value of the charge mode. This is done by adding the following perturbation:
\begin{align}
&g_{K\perp B}'\int dx\cos\sqrt{2}\phi_{1c}\cos 2\theta_-\notag\\
+&g_{KzB+}'\int dx\cos\sqrt{2}\phi_{1c}\cos 2\phi_+
\label{ptb_SPT_bos}
\end{align}
which is generated by an artificial Kondo coupling
\begin{equation}
H'_K=J'\sum_{j,\sigma,\sigma'}c_{j\sigma}^\dag\bm{\sigma}_{\sigma\sigma'}c_{j+1\sigma'}\cdot \bm{S}_{fj}+\mathrm{h.c.}.
\label{ptb_SPT}
\end{equation}
The shift of the expectation value of $\phi_{1c}$ by $\pi$ is equivalent to the sign reversal of the Kondo couplings $H_{K\perp B}, H_{KzB+}$, and $H_{KzB-}$, and thus this procedure connects the topological phase with the trivial Kondo insulator.
As seen from Eq.\ \eqref{ptb_SPT} or Table \ref{SymTfm}, this perturbation only breaks the inversion symmetry, and preserves the other symmetries. In the present system, the charge U(1) symmetry prohibits vertex operators which involve the field $\theta_{1c}$. 
Thus, the only possible way to connect the topological phase and the trivial phase using the charge degrees of freedom is the shift of the expectation value of $\phi_{1c}$ accompanied by the breaking of inversion symmetry. From these observations, we conclude that the topological phase is protected only by the inversion symmetry (under the assumption of the charge conservation).

From the above argument, we can interpret the crossover from the fermionic SPT phase (protected by the inversion symmetry only) to the bosonic SPT phase (the Haldane phase, protected by the time-reversal, inversion, and spin dihedral symmetries) via the bosonization language. In the weakly interacting regime, the low-energy behavior of the charge mode is mainly determined by the Kondo coupling rather than the Umklapp scattering due to the Hubbard repulsion. In this case, we can connect the topological phase and the trivial phase by shifting the pinning position of the charge mode with breaking the inversion symmetry, while the time-reversal and the spin rotation symmetries are kept intact. However, this shift cannot be reconciled with minimization of the Umklapp term $H_{U}$. Hence if we gradually increase the Hubbard repulsion $U$, the above procedure fails to work at some point. After that, the topological phase and the trivial phase are separated by a quantum phase transition if the time-reversal or the spin dihedral symmetry is present. We note that the perturbation \eqref{ptb_SPT_bos} vanishes if the charge mode is frozen at the potential minimum of the Umklapp term, $2\sqrt{2}\phi_{1c}=\pi$.

Finally, let us clarify where the topological phase of the Kondo lattice stands in the classification of SPT phases of interacting fermions. In non-interacting systems, topological insulators protected by the inversion symmetry in 1D are classified\cite{ShiozakiSato, LuLee, Hughes} by integer $\mathbb{Z}$, which means that there are infinitely many different topological phases. However, when we allow interactions as perturbation to systems, a part of nontrivial topological phases can be connected to the trivial phase and free-fermion classification of topological phases is reduced to its subgroup.\cite{FidkowskiKitaev1, FidkowskiKitaev2, Turner, Morimoto, Kapustin} In the case of inversion-symmetric topological insulators, the classification is performed by several methods\cite{YouXu, Shiozaki1, Shiozaki2} and is argued to reduce from $\mathbb{Z}$ to $\mathbb{Z}_4$ in the interacting case. Since the Haldane phase is classified by $\mathbb{Z}_2$, two copies of them can be deformed into the trivial phase. Using the fact that the topological phase of the 1D Kondo lattice approaches the Haldane phase in the strong coupling limit, we can also deform the two copies of the model \eqref{H_bos} into a trivial phase. Thus we conclude that the topological phase in 1D Kondo lattice is specified by an integer $2\in\mathbb{Z}_4=\{0,1,2,3\}$.

\begin{center}
\begin{table*}
\caption{Symmetry transformation in bosonization. The transformation on boson fields $\phi_{2s}, \theta_{2s}$ is the same as that on $\phi_{1s}, \theta_{1s}$ in the table.\label{SymTfm}}
\begin{ruledtabular}
\begin{tabular}{lcccc}
Symmetry operation & \multicolumn{2}{c}{Transformation law} & \multicolumn{2}{c}{Transformation on boson fields} \\ \hline
Translation & $c_\sigma(x)\to c_\sigma(x+a),$ & $\bm{S}_f(x)\to \bm{S}_f(x+a)$ & $\phi_{1c}(x)\to\phi_{1c}(x+a)-\sqrt{2}k_F a,$ & $\theta_{1c}(x)\to\theta_{1c}(x+a)$ \\
& & & $\phi_{1s}(x)\to\phi_{1s}(x+a),$ & $\theta_{1s}(x)\to\theta_{1s}(x+a)$\\
Charge U(1) & $c_\sigma\to e^{i\varphi}c_\sigma$ & & $\phi_{1c}\to\phi_{1c},$ & $\theta_{1c}\to\theta_{1c}+\varphi$\\
& & & $\phi_{1s}\to\phi_{1s},$ & $\theta_{1s}\to\theta_{1s}$\\
Time reversal & $c_\sigma\to \sum_{\sigma'}(i\sigma_y)_{\sigma\sigma'}c_{\sigma'},$ & $\bm{S}_f\to -\bm{S}_f$ & $\phi_{1c}\to\phi_{1c},$ & $\theta_{1c}\to -\theta_{1c}+\frac{\pi}{\sqrt{2}}$ \\
& & & $\phi_{1s}\to -\phi_{1s},$ & $\theta_{1s}\to\theta_{1s}-\frac{\pi}{\sqrt{2}}$ \\
Spatial inversion & $c_\sigma(x)\to c_{\sigma}(a-x),$ & $\bm{S}_f(x)\to\bm{S}_f(a-x)$ & $\phi_{1c}(x)\to -\phi_{1c}(a-x)+\sqrt{2}k_F a,$ & $\theta_{1c}(x)\to\theta_{1c}(a-x)$ \\
& & & $\phi_{1s}(x)\to -\phi_{1s}(a-x),$ & $\theta_{1s}(x)\to\theta_{1s}(a-x)$ \\
$\pi$ rotation around $x$ axis & \multicolumn{2}{c}{$S_{c,f}^x\to S_{c,f}^x,S_{c,f}^{y,z}\to -S_{c,f}^{y,z}$} & $\phi_{1c}\to\phi_{1c},$ & $\theta_{1c}\to\theta_{1c}$ \\
& & & $\phi_{1s}\to -\phi_{1s},$ & $\theta_{1s}\to -\theta_{1s}$ \\
$\pi$ rotation around $y$ axis & \multicolumn{2}{c}{$S_{c,f}^y\to S_{c,f}^y,S_{c,f}^{x,z}\to -S_{c,f}^{x,z}$} & $\phi_{1c}\to\phi_{1c},$ & $\theta_{1c}\to\theta_{1c}$ \\
& & & $\phi_{1s}\to -\phi_{1s},$ & $\theta_{1s}\to -\theta_{1s}+\frac{\pi}{\sqrt{2}}$ \\
Spin U(1) & \multicolumn{2}{c}{$S_{c,f}^x\to S_{c,f}^x\cos\varphi+S_{c,f}^y\sin\varphi,$} & $\phi_{1c}\to\phi_{1c},$ & $\theta_{1c}\to\theta_{1c}$ \\
& \multicolumn{2}{c}{$S_{c,f}^y\to -S_{c,f}^x\sin\varphi+S_{c,f}^y\cos\varphi$} & $\phi_{1s}\to \phi_{1s},$ & $\theta_{1s}\to \theta_{1s}+\varphi$ \\
\end{tabular}
\end{ruledtabular}
\end{table*}
\end{center}

\subsection{Protection by spin $\pi$ rotation symmetries around the $x$ or $y$ axis}
Besides the topological protection described in the previous subsection, the gapped phases in this system are also protected by spin $\pi$ rotation symmetries around the $x$ or $y$ axis. This gives a distinction between the laser-induced Kondo insulator (K') with the ``unusual" spin state and the two phases (Top) and (K) composed of the ordinary Kondo singlet state. Although this fact is not related to the discussion of SPT phases, we describe the mechanism of this symmetry protection for completeness. This fact arises from the symmetry eigenvalues of the spin $\pi$ rotation symmetries. In the description of SPT phases using matrix product states, the symmetry eigenvalues correspond to phase factors which are not related to topological phases and provide distinction between ``trivial" phases (see Appendix \ref{App_SPT}). 
To calculate the phase factors, we use a strong coupling limit $|J_\perp|, |J_z|\to\infty$, since the phase factors cannot change unless the energy gap collapses. The strong coupling limit of the topological phase is continuously connected to the Haldane phase of the spin-1 Heisenberg model, and therefore we obtain $\vartheta_{x}=\vartheta_y=0$ (see the notation in Appendix \ref{App_SPT}). In the strong coupling limit of the (ordinary) Kondo insulator and the laser-induced Kondo insulator, the ground states are site-product states of on-site Kondo singlets. The Kondo singlet is $\ket{\down}_c\ket{\up}_f-\ket{\up}_c\ket{\down}_f$ for the former phase, and $\ket{\down}_c\ket{\up}_f+\ket{\up}_c\ket{\down}_f$ for the latter phase, respectively. Since the spin $\pi$ rotational operation $R_x$ around the $x$ axis satisfies
\begin{align}
R_{x}(\ket{\down}_c\ket{\up}_f-\ket{\up}_c\ket{\down}_f)&=+(\ket{\down}_c\ket{\up}_f-\ket{\up}_c\ket{\down}_f),\\
R_{x}(\ket{\down}_c\ket{\up}_f+\ket{\up}_c\ket{\down}_f)&=-(\ket{\down}_c\ket{\up}_f+\ket{\up}_c\ket{\down}_f),
\end{align}
and the same holds for $R_y$, we obtain $\vartheta_{x}=\vartheta_y=0$ for the ordinary Kondo insulator and $\vartheta_{x}=\vartheta_y=\pi$ for the laser-induced Kondo insulator. By comparing $\vartheta_{x}, \vartheta_y$ of each phase, we conclude that the laser-induced Kondo insulating phase (K') is distinct from the ordinary Kondo insulator (K) and the topological phase (Top), protected by the spin $\pi$ rotation symmetry around the $x$ or $y$ axis. To connect the distinct phases, we must close the energy gap or break the symmetry. In fact, at the phase boundary between the topological phase and the laser-induced Kondo insulator, the spin gap of $\phi_+$ is collapsed. At the boundary between the ordinary and the laser-induced Kondo insulators, the N\'{e}el order intervenes, signaling the symmetry breaking. Thus, the phase diagram obtained in Sec.\ \ref{Phasediagram} is consistent with the symmetry protection. 

If the spin $\pi$ rotation symmetries are broken, we can adiabatically connect the laser-induced Kondo insulator and the ordinary Kondo insulator. To check this, let us consider a unitary transformation\cite{MaruyamaHatsugai} $U(\gamma)^\dag H U(\gamma)$ with
\begin{equation}
U(\gamma)=\exp\Bigl[i\gamma\sum_j S_{fj}^z\Bigr],
\end{equation}
which changes the Kondo coupling into
\begin{align}
&U(\gamma)^\dag H_K U(\gamma)\notag\\
=&-J_\perp\cos\gamma\sum_j (S_{cj}^x S_{fj}^x+S_{cj}^y S_{fj}^y)-J_z\sum_j S_{cj}^z S_{fj}^z\notag\\
&-J_\perp\sin\gamma\sum_j(S_{cj}^x S_{fj}^y-S_{cj}^y S_{fj}^x).
\label{tfm_SPt}
\end{align}
The rest of the Hamiltonian is unchanged. As seen easily, the spin $\pi$ rotation symmetry around the $x$ or $y$ axis is broken in the transformed Hamiltonian except for $\gamma=0,\pi$. Since $U(\gamma)$ is unitary, the energy spectra of $H$ and $U(\gamma)^\dag H U(\gamma)$ are identical. Thus we can connect the ordinary Kondo insulator at $\gamma=0$ and the laser-induced Kondo insulator at $\gamma=\pi$ without closing the energy gap by changing $\gamma$ continuously.

We can also show the symmetry protection using the bosonization language. Let us focus on a parameter region near the phase boundary between the topological phase and the laser-induced Kondo insulator. In that region, the relevant perturbation for the gap generation in terms of the scaling dimensions is $H_{K\perp B}$ and $H_{KzB+}$, and the low-energy behavior is governed by these terms, making the fields $\phi_{1c}, \phi_+$, and $\theta_-$ locked at their potential minimum. Here we note that the $H_{K\perp B}$ term does not change its sign between the two phases, but the $H_{KzB+}$ term does. Hence the difference between the two phases is the pinning position of the total spin mode $\phi_+$. To adiabatically connect the two phases preserving the energy gap, we must shift the expectation value of $\phi_+$ by allowing a perturbation term like
\begin{equation}
g_{KzB+}'\int dx \sin\sqrt{2}\phi_{1c}\sin2\phi_+.
\label{ptb_SPt}
\end{equation}
We note that in the present system an additional spin U(1) symmetry forbids perturbations containing the dual field $\theta_+$. However, the shift of the expectation value of $\phi_+$ necessarily breaks the spin $\pi$ rotation symmetry as inferred from Table \ref{SymTfm}. 
Hence the quantum phase transition between the topological phase and the laser-induced Kondo insulator is protected by the spin $\pi$ rotation symmetry, being consistent with the analysis of symmetry eigenvalues.

To connect the ordinary Kondo insulator and the laser-induced Kondo insulator, we must shift the expectation value of $\theta_-$. This procedure also breaks the spin $\pi$ rotation symmetries. The required perturbation can be obtained by bosonization of the last term in Eq.\ \eqref{tfm_SPt}.

\section{Discussions and Conclusion\label{Conclusion}}
We have shown that cold-atom realization of the Kondo lattice model offers a platform to investigate a 1D SPT phase and an associated quantum phase transition with high controllability. By utilizing the spin-exchanging collisions with the help of the laser-induced mixing of internal states, ultracold AEA in optical lattice can realize the Kondo lattice with tunable anisotropic Kondo couplings, which is hard to be realized in solid state experiments. Since the sign of the bare exchange coupling $V_{\mathrm{ex}}$ can be controlled using the confinement-induced resonance specific to 1D optical lattices,\cite{RZhang} a large portion of the phase diagram in Fig.\ \ref{fig_phase} can be accessed in this system. If we start from ferromagnetic $V_{\mathrm{ex}}>0$, the SPT phase transition from the topological phase to the laser-induced Kondo insulating state is possible. This phase transition is protected by the inversion symmetry and the spin $\pi$ rotation symmetries around the $x$ or $y$ axis, and the only former symmetry stands for the topological properties. On the other hand, if we switch on the laser coupling starting from antiferromagnetic $V_{\mathrm{ex}}<0$, the ordinary Kondo insulator is first changed into the N\'{e}el order, and finally turns into the laser-induced Kondo insulator. This reentrant Kondo transitions associated with the N\'{e}el order are stable (at least $T=0$) if the spin $\pi$ rotation symmetries are preserved.

We have also demonstrated the topological phase of the 1D Kondo lattice is protected only by the inversion symmetry when the charge fluctuations cannot be ignored, while the Haldane phase in the strong coupling limit is also protected by the time-reversal and spin dihedral symmetries. The change of the nature of the topological phase from fermionic to bosonic SPT phases leads to an intriguing consequence in the fate of edge states of the topological phase. In the strong coupling regime, the Haldane phase has spin-1/2 zero-energy states at the edge of the system. The edge states are magnetically active, and have been detected by applying magnetic fields.\cite{Hagiwara,Glarum} On the other hand, in the weak-coupling regime, the SPT phase is protected only by the inversion symmetry. This means that the zero-energy edge state is absent in general, since the edges generically break the inversion symmetry. Thus it is implied that the edge states gradually decrease their excitation energies with increasing the Hubbard interaction $U$, and finally they turn into the zero-energy state at some threshold value of $U$. Such ``interaction-induced" edge states are one possible hallmark of the crossover from fermionic SPT phases to bosonic ones.

Observation of such a crossover using the present cold-atom setup is intriguing but may be a challenging issue. To detect a clear signature of the edge states, it is appropriate to create an interface between the topologically nontrivial phase and the trivial phase,\cite{Leder, Goldman} since the true edge of the atomic cloud is usually a metallic state due to a harmonic confinement potential. In our setup, the interface can be easily created, since the topological-trivial phase transition is caused by the laser irradiation, which can be performed in a spatially varying manner. The interface-localized edge modes are, in principle, detected by combining a magnetic field and spin-resolved quantum gas microscopy, by which antiferromagnetic correlations were recently observed in the Fermi-Hubbard model.\cite{Parsons, Boll, Cheuk, Mazurenko}

\begin{acknowledgments}
We are grateful to Ken Shiozaki for useful discussions, and acknowledge Tsuneya Yoshida, Takahiro Morimoto, and Akira Furusaki for helpful comments at the early stage of this work. This work was supported by JSPS KAKENHI (Grants No.\ JP16K05501 and No.\ JP14J01328) and a Grand-in-Aid for Scientific Research on Innovative Areas (Grant No.\ JP15H05855). M.\ N.\ was supported by a JSPS Research Fellowship for Young Scientists and RIKEN Special Postdoctoral Researcher Program.
\end{acknowledgments}

\appendix
\section{Bosonization of 1D Kondo lattice\label{App_bos}}
In this Appendix, we derive the bosonized Hamiltonian \eqref{H_bos}-\eqref{Hint} from the model \eqref{Heff}. We divide the Hamiltonian into three parts:
\begin{align}
H_{\mathrm{eff}}&=H_c+H_f+H_K,\label{H}\\
H_c&=-t_c\sum_{j,\sigma}(c_{j\sigma}^\dag c_{j+1\sigma}+\mathrm{h.c.})+U\sum_j n_{cj\up}n_{cj\down},\label{Hc}\\
H_f&=J_H\sum_j\bm{S}_{fj}\cdot\bm{S}_{fj+1},\label{Hf}\\
H_K&=-J_\perp\sum_j (S_{cj}^x S_{fj}^x+S_{cj}^y S_{fj}^y)-J_z\sum_j S_{cj}^z S_{fj}^z.\label{HK}
\end{align}
To apply the bosonization recipe, we focus on the low-energy behavior of the system and linearize the dispersion relation of the Hubbard part \eqref{Hc}. Then Eq.\ \eqref{Hc} is bosonized as
\begin{align}
H_c=&\sum_{\nu=c,s}\frac{1}{2\pi}\int dx(u_{1\nu}K_{1\nu}(\nabla\theta_{1\nu})^2+\frac{u_{1\nu}}{K_{1\nu}}(\nabla\phi_{1\nu})^2)\notag\\
&+\frac{U}{2\pi^2\alpha}\int dx \cos(2\sqrt{2}\phi_{1c}),
\label{Hubbard_bos}
\end{align}
where a marginally irrelevant term in the spin part is neglected. $\alpha$ denotes a short-range cutoff. Here the charge mode and the spin mode are defined as $\phi_{1c,1s}=\frac{1}{\sqrt{2}}(\phi_{1\up}\pm\phi_{1\down}), \theta_{1c,1s}=\frac{1}{\sqrt{2}}(\theta_{1\up}\pm\theta_{1\down})$, respectively (the minus sign stands for the spin part). The cosine term comes from the Umklapp scattering due to the Hubbard interaction. The velocities are $u_{1c}=2t_c a\sqrt{1+\frac{U}{2\pi t_c}}, u_{1s}=2t_c a\sqrt{1-\frac{U}{2\pi t_c}}$ and the Luttinger parameters are $K_{1c}=1/\sqrt{1+\frac{U}{2\pi t_c}}, K_{1s}=1$. The Luttinger parameter for the spin part has been set unity because of the spin SU(2) symmetry of the Hubbard part.

The Heisenberg part \eqref{Hf} is also bosonized. While one can use the standard Jordan-Wigner transformation to convert the spin chain into fermions, we here adopt an expression of the Heisenberg chain as the Mott insulating phase of the Hubbard model, where the charge mode is gapped out by the cosine term in Eq.\ \eqref{Hubbard_bos}, since a parallel description is available between the $^1S_0$ and the $^3P_0$ states. Then the Heisenberg chain is described by the spin part of the bosonized Hubbard Hamiltonian as
\begin{align}
H_f=\frac{1}{2\pi}\int dx(u_{2}(\nabla\theta_{2s})^2+u_{2}(\nabla\phi_{2s})^2)
\label{Heis_bos}
\end{align}
where we again set the Luttinger parameter as unity due to the SU(2) symmetry. For simplicity, we assume $u_{1s}=u_2\equiv u$.

Finally, we bosonize the Kondo coupling \eqref{HK}. The spin operators of the $^1S_0$ state are expressed as
\begin{align}
\bm{S}_c(x)\equiv\bm{S}_{cj}/\alpha=\bm{M}_c(x)+(-1)^{x/a}\bm{N}_c(x).
\end{align}
The uniform component $\bm{M}_c(x)$ reads
\begin{subequations}
\begin{align}
M_{c}^x(x)=&\frac{1}{\pi\alpha}
\sin\sqrt{2}\theta_{1s}\cos\sqrt{2}\phi_{1s},\label{Mcx}\\
M_{c}^y(x)=&\frac{1}{\pi\alpha}
\cos\sqrt{2}\theta_{1s}\cos\sqrt{2}\phi_{1s},\label{Mcy}\\
M_c^z(x)=&-\frac{1}{\sqrt{2}\pi}\nabla\phi_{1s}\label{Mcz}
\end{align}
\end{subequations}
and the staggered component $\bm{N}_c(x)$ is
\begin{subequations}
\begin{align}
N_{c}^x(x)=&\frac{1}{\pi\alpha}
\cos\sqrt{2}\theta_{1s}\sin\sqrt{2}\phi_{1c},\label{Ncx}\\
N_{c}^y(x)=&-\frac{1}{\pi\alpha}
\sin\sqrt{2}\theta_{1s}\sin\sqrt{2}\phi_{1c},\label{Ncy}\\
N_c^z(x)=&\frac{1}{\pi\alpha}
\cos\sqrt{2}\phi_{1s}\sin\sqrt{2}\phi_{1c}\label{Ncz}.
\end{align}
\end{subequations}
Those of the $^3P_0$ state, $\bm{M}_f(x)$ and $\bm{N}_f(x)$, are of the same form as Eqs.\ \eqref{Mcx} - \eqref{Mcz} and \eqref{Ncx} - \eqref{Ncz} with the charge mode replaced by its expectation value $m=\langle\sin\sqrt{2}\phi_{2c}\rangle$. The Kondo coupling $H_K$ is thereby divided into the following parts:
\begin{align}
H_{K\perp F}=&-J_\perp\int dx(M_c^x(x)M_f^x(x)+M_c^y(x)M_f^y(x))\notag\\
=&H_{K\perp F+}+H_{K\perp F-},\\
H_{K\perp B}=&-J_\perp\int dx(N_c^x(x)N_f^x(x)+N_c^y(x)N_f^y(x))\notag\\
=&-g_{K\perp B}
\int dx\sin\sqrt{2}\phi_{1c}\cos 2\theta_-,\\
H_{KzF}=&-J_z\int dx M_c^z(x)M_f^z(x)\notag\\
=&-\frac{\alpha J_z}{2\pi^2}\int dx\nabla\phi_{1s}\nabla\phi_2,\label{HKzF}\\
H_{KzB}=&-J_z\int dx N_c^z(x)N_f^z(x)\notag\\
=&H_{KzB+}+H_{KzB-}
\end{align}
with
\begin{align}
H_{K\perp F+}=&-g_{K\perp F+}
\int dx\cos 2\phi_+\cos 2\theta_-,\\
H_{K\perp F-}=&-g_{K\perp F-}
\int dx\cos 2\phi_-\cos 2\theta_-,\\
H_{KzB+}=&-g_{KzB+}
\int dx\sin\sqrt{2}\phi_{1c}\cos 2\phi_+,\\
H_{KzB-}=&-g_{KzB-}
\int dx\sin\sqrt{2}\phi_{1c}\cos 2\phi_-,
\end{align}
where the coupling constants are
\begin{gather}
g_{K\perp F+}=g_{K\perp F-}=\frac{1}{2m}g_{K\perp B}=\frac{J_\perp}{2\pi^2\alpha},\\
g_{KzB+}=g_{KzB-}=\frac{mJ_z}{2\pi^2\alpha}.
\end{gather}
Here we have named each perturbation with subscript $F$ and $B$ in terms of momentum transfers of itinerant fermions in analogy with the impurity problem. In the calculation, we have dropped the oscillation terms which vanish after the integration. Also we have defined new boson fields $\phi_\pm, \theta_\pm$ as
\begin{align}
\phi_\pm\equiv\frac{1}{\sqrt{2}}(\phi_{1s}\pm\phi_{2s}),\\
\theta_\pm\equiv\frac{1}{\sqrt{2}}(\theta_{1s}\pm\theta_{2s}),
\end{align}
which describe the total ($+$) and relative ($-$) spin modes, respectively. After combining the quadratic \eqref{HKzF} term into the non-interacting parts, we obtain the bosonized Hamiltonian \eqref{H_bos}-\eqref{Hint}.

\section{SPT phase in one dimension\label{App_SPT}}
In this appendix, we briefly review the basic properties of 1D SPT phases\cite{Chen2, Chen3, GuWen, Pollmann1, Pollmann2} used in the main text. SPT phases in one spatial dimension are well understood thanks to a universal description of gapped ground states using matrix product states (MPS).\cite{Chen2, Chen3} A generic 1D ground state can be described by MPS in a canonical form as
\begin{equation}
\ket{\psi}=\sum_{\{ i_n\}}\mathrm{Tr}[\Gamma_{i_1}\Lambda\Gamma_{i_2}\Lambda\cdots\Gamma_{i_L}]\ket{i_1}\otimes\ket{i_2}\otimes\cdots\otimes\ket{i_L}
\label{MPS}
\end{equation}
where we denote the number of sites as $L$. For simplicity, here we assume the periodic boundary condition and translational invariance. $\{\ket{i}\}_i$ is the basis of the Hilbert space at each site. $\Gamma_i$ is a square matrix and the elements of the diagonal matrix $\Lambda$ are related to the entanglement spectrum. The symmetry protection of the Haldane phase in spin chains can be proved by using MPS. Let us consider the following three symmetry operations: the time-reversal ($T$), spatial inversion ($I$), and the spin dihedral symmetry composed of $\pi$ rotation around each axis ($R_x, R_y, R_z$). When the on-site basis transforms as $\ket{i}\to\sum_j(\tau_g)_{ji}\ket{j}$ by a symmetry operation $g$, $\Gamma_i$ transforms as
\begin{align}
\sum_j (\tau_T)_{ij}(\Gamma_j)^*&=e^{i\vartheta_T}U_T^\dag\Gamma_i U_T,\\
U_TU_T^*&=e^{i\varphi_T}.
\end{align}
for time reversal. Similarly, for spatial inversion,
\begin{align}
(\Gamma_i)^T&=e^{i\vartheta_I}U_I^\dag\Gamma_i U_I,\\
U_IU_I^*&=e^{i\varphi_I}.
\end{align}
For the spin $\pi$ rotation around the $\alpha=x,y,z$ axis,
\begin{align}
\sum_j (\tau_\alpha)_{ij}\Gamma_j&=e^{i\vartheta_\alpha}U_\alpha^\dag\Gamma_i U_\alpha,\\
U_xU_z&=e^{i\varphi_{xz}}U_zU_x.
\end{align}
$U_T, U_I, U_\alpha$ are unitary matrices, and the phase factors $\vartheta_I, \vartheta_\alpha$ and $\varphi_T, \varphi_I, \varphi_{xz}$ are quantized\cite{memo_thetaT} to $0$ or $\pi$. Since the discrete phase factors cannot be changed unless the symmetries are broken or the energy gap collapses, ground states which have different phase factors are necessarily separated by quantum phase transitions. Among the phase factors, the quantization of $\varphi_T, \varphi_I, \varphi_{xz}$ leads to the nontrivial SPT phase (the Haldane phase), which is reflected in the degeneracy of the entanglement spectrum.\cite{Pollmann1} In contrast, the phase factors $\vartheta_I, \vartheta_\alpha$ mean the eigenvalues of the ground state under symmetry operations, since the symmetry operation maps the ground state $\ket{\psi}$ into $e^{iL\vartheta_\alpha}\ket{\psi}$ $(\alpha=I, x,y, z)$. In this case, the quantization simply comes from the property $I^2=R_x^2=R_y^2=R_z^2=1$. Thus, although the quantization can distinguish quantum phases which have different symmetry eigenvalues, this property does not lead to SPT phases. In this sense, the quantization of $\vartheta_I, \vartheta_{x,y,z}$ diagnoses the distinction between trivial phases. Under certain point-group symmetry, quantization of a combination of $\vartheta_\alpha$ and $\varphi_\alpha$ can also lead to distinct trivial phases.\cite{FujiPollmannOshikawa} For the Haldane phase, we can calculate the phase factors using the exact ground state (Affleck-Kennedy-Lieb-Tasaki state\cite{AKLT1, AKLT2}) as\cite{Pollmann1, Pollmann2} $\vartheta_I=\pi, \vartheta_x=\vartheta_y=\vartheta_z=0, \varphi_T=\varphi_I=\varphi_{xz}=\pi$. Hence, the Haldane phase is stable if either of the three symmetries is present. 

The structure of symmetry protection of SPT phases should be encoded in their low-energy effective theory. Description of symmetry protection in bosonization language was discussed in Refs.\ \onlinecite{Berg, Fuji}. To exemplify the symmetry protection using bosonization approach, let us consider the following sine-Gordon theory:
\begin{equation}
H=\frac{1}{2\pi}\int dx(uK(\nabla\theta)^2+\frac{u}{K}(\nabla\phi)^2+g\cos\frac{\phi}{R}).
\end{equation}
We here assume that the boson field $\phi$ is compactified with radius $R$ (namely $\phi\sim\phi+2\pi R$), and hence 
the cosine term appearing in the Hamiltonian is the most relevant perturbation allowed in the system. For simplicity, we do not consider a vertex operator which contains the dual field $\theta$ or simply assume that such a term is forbidden by a symmetry. The ground state of this theory is gapped if $K<8R^2$, and the boson field $\phi$ is pinned at $\phi=0$ for $g<0$ and $\phi=\pi R$ for $g>0$. The two phases are separated by a critical point at $g=0$. If the sine term $\sin\frac{\phi}{R}$ is forbidden by a symmetry constraint, we can say that the two phases cannot be adiabatically connected, since the cosine term is the most relevant perturbation in the system and the critical point cannot be gapped without breaking the symmetry. Conversely, we can connect the two phases if the sine term is allowed, since $g\cos\frac{\phi}{R}+g'\sin\frac{\phi}{R}=G\cos(\frac{\phi}{R}+\gamma)$ and the parameter $\gamma$ can be changed from $0$ to $\pi$ by tuning the ratio between $g$ and $g'$.

\bibliography{KondoHaldane_ref.bib}

\begin{thebibliography}{102}
\bibitem [{mem()}]{memo_thetaT}%
  \BibitemOpen
  \href@noop {} {}\bibinfo {note} {$\vartheta_T$ is not well-defined since it
  depends on the choice of the phase of the ground
  state\cite{Pollmann1}}\BibitemShut {NoStop}%
\end{thebibliography}

\end{document}